\documentclass[aps,prb,reprint,superscriptaddress,longbibliography,nofootinbib]{revtex4-2}

\usepackage{graphicx,amsmath,amssymb,amsfonts,dcolumn,dsfont,latexsym,color,bm,hyperref,siunitx,mathrsfs,braket}
\usepackage[utf8]{inputenc}  

\DeclareUnicodeCharacter{0301}{\'{}} 

\hypersetup{
    colorlinks=true,
    linkcolor=blue,
    citecolor=blue,
    urlcolor=blue,
}

\begin{document}

\title{Diode effect in a skyrmion-coupled high-temperature Josephson junction}

\author{Digvijay Singh}
\email{digvijay\_s@ph.iitr.ac.in}
\affiliation{Department of Physics, Indian Institute of Technology Roorkee, Roorkee 247667, India}

\author{Pankaj Sharma}
\email{pankaj@ph.iitr.ac.in}
\affiliation{Department of Physics, Indian Institute of Technology Roorkee, Roorkee 247667, India}

\thanks{These authors contributed equally to this work.}

\author{Narayan Mohanta}
\email{narayan.mohanta@ph.iitr.ac.in}
\affiliation{Department of Physics, Indian Institute of Technology Roorkee, Roorkee 247667, India}

\begingroup
\renewcommand\thefootnote{} 
\footnotetext{D.S. and P.S. contributed equally to this work.}
\endgroup

\begin{abstract}
We show that a planar Josephson junction having $d$-wave superconducting regions, with a skyrmion crystal placed underneath, produces a robust gate-tunable superconducting diode effect. The spatially-varying exchange field of the skyrmion crystal breaks both inversion and time-reversal symmetries, leading to an asymmetric current-phase relation with an anomalous phase shift. Our theoretical calculations, obtained using resistively and capacitively shunted junction model combined with Bogoliubov-de Gennes method, reveal that the diode efficiency is largely tunable by controlling external gate voltage and skyrmion radius. Incorporation of a $d$-wave superconductor such as high-$T_c$ Cuprate enables the diode to function at higher operating temperatures. Our results establish a unique and practically-realizable mechanism for devising tunable field-free superconducting diodes based on magnetic texture-superconductor hybrid platforms.
\end{abstract}

\maketitle

\section{Introduction}

The discovery of nonreciprocal superconducting transport has opened a new frontier in nonequilibrium quantum phenomena, providing a superconducting analogue of the conventional semiconductor diode. 
In such systems, the maximum dissipationless current that can flow---the critical supercurrent---depends on the direction of current flow, $I_c^{+} \neq |I_c^{-}|$, thereby enabling rectification of supercurrents without energy dissipation. 
This phenomenon, known as the \emph{superconducting diode effect} or, in Josephson systems, the \emph{Josephson diode effect}, has recently been demonstrated across a wide variety of material platforms~\cite{ Ando_Nature_2020, Baumgartner_Nat.Nanotechnol._2022, Daido_Phys.Rev.Lett._2022, Davydova_Sci.Adv._2022, Diez-Merida_NatCommun_2023, Golod_NatCommun_2022, Gutfreund_NatCommun_2023, Haim_Phys.Rev.B_2019, He_NewJ.Phys._2022, Hou_Phys.Rev.Lett._2023, Ilic_Phys.Rev.Lett._2022, Jeon_Nat.Mater._2022, Ke_NatCommun_2019, Kleiner_Proc.IEEE_2004, Kuiri_NatCommun_2022, Lin_Nat.Phys._2022, Hess_PRB_2023, Liu_Sci.Adv._2024, Ma_CommunPhys_2025, Misaki_Phys.Rev.B_2021, Narita_Nat.Nanotechnol._2022, Pal_Nat.Phys._2022, Reinhardt_NatCommun_2024, Seshadri_SciPostPhys._2022, Steiner_Phys.Rev.Lett._2023, Strambini_NatCommun_2022, Volkov_Phys.Rev.B_2024, Wu_Nature_2022, Yokoyama_Phys.Rev.B_2015, Yuan_Proc.Natl.Acad.Sci._2022, Zhao_Science_2023,Vakili_Phys.Rev.B_2024, Trahms_Nature_2023}. 
The diode effect has been reported in noncentrosymmetric thin films~\cite{Hou_Phys.Rev.Lett._2023}, van der Waals heterostructures~\cite{Wu_Nature_2022,Ma_CommunPhys_2025}, topological semimetals~\cite{Pal_Nat.Phys._2022}, twisted graphene systems~\cite{Lin_Nat.Phys._2022,Diez-Merida_NatCommun_2023}, and ferromagnet–superconductor multilayers~\cite{Narita_Nat.Nanotechnol._2022}. 
Concurrently, theoretical studies have elucidated several distinct microscopic mechanisms for superconducting diode effect and Josephson diode effect, including spin--orbit coupling induced magnetochiral anisotropy~\cite{Baumgartner_Nat.Nanotechnol._2022}, finite--momentum Cooper pairing~\cite{Yuan_Proc.Natl.Acad.Sci._2022,Davydova_Sci.Adv._2022}, and time-reversal symmetry breaking in systems with noncentrosymmetric order parameters~\cite{Daido_Phys.Rev.Lett._2022}. 
Together, these advances have established that the coexistence of broken inversion and time--reversal symmetries is a fundamental prerequisite for achieving nonreciprocal superconductivity.

Most known realizations of the Josephson diode effect rely on uniform Zeeman fields or interfacial Rashba coupling to achieve time-reversal and inversion symmetry breaking. Recent developments in spintronic and topological materials have revealed new possibilities for generating effective magnetic fields and spin--orbit coupling internally through \emph{real-space magnetic textures}. 
In particular, magnetic skyrmions (topologically protected spin configurations characterized by a finite winding number) naturally break both inversion and time-reversal symmetries through their noncollinear and noncoplanar spin structure. 
When coupled to superconductors, skyrmions can induce emergent electromagnetic fields and spin-triplet correlations that profoundly modify quasiparticle dynamics~\cite{Yokoyama_Phys.Rev.B_2015,Mohanta_CommunPhys_2021,Hess_PRB_2023}. 
These features make skyrmion–superconductor hybrids an attractive platform for realizing new types of Josephson phenomena, including anomalous phase shifts and nonreciprocal supercurrents.

Planar Josephson junctions have recently emerged as a particularly versatile architecture for studying both conventional and topological superconductivity. 
Unlike nanowire--based devices, planar Josephson junction offer continuous control over key parameters such as carrier density, junction width, and phase bias via electrostatic gating~\cite{Pientka_Phys.Rev.X_2017,Ren_Nature_2019,Banerjee_Phys.Rev.B_2023,Schiela_PRXQuantum_2024,Sharma_Phys.Rev.B_2024, Sharma_Phys.Rev.B_2025, Sharma__2025}. 
They have enabled the realization of gate--tunable $\varphi_0$ junctions~\cite{Reinhardt_NatCommun_2024} and current--biased diode effects~\cite{Steiner_Phys.Rev.Lett._2023}. 
The flexibility of this architecture extends naturally to integrating complex magnetic textures, such as skyrmion lattices, beneath the superconducting leads. 
The coupling between the momentum--space anisotropy of unconventional pairing (e.g., $d$--wave) and the real--space topology of skyrmion spin textures offers an unexplored route toward field--free superconducting diode behavior.

\begin{figure}[t]
\includegraphics[width=0.9\linewidth]{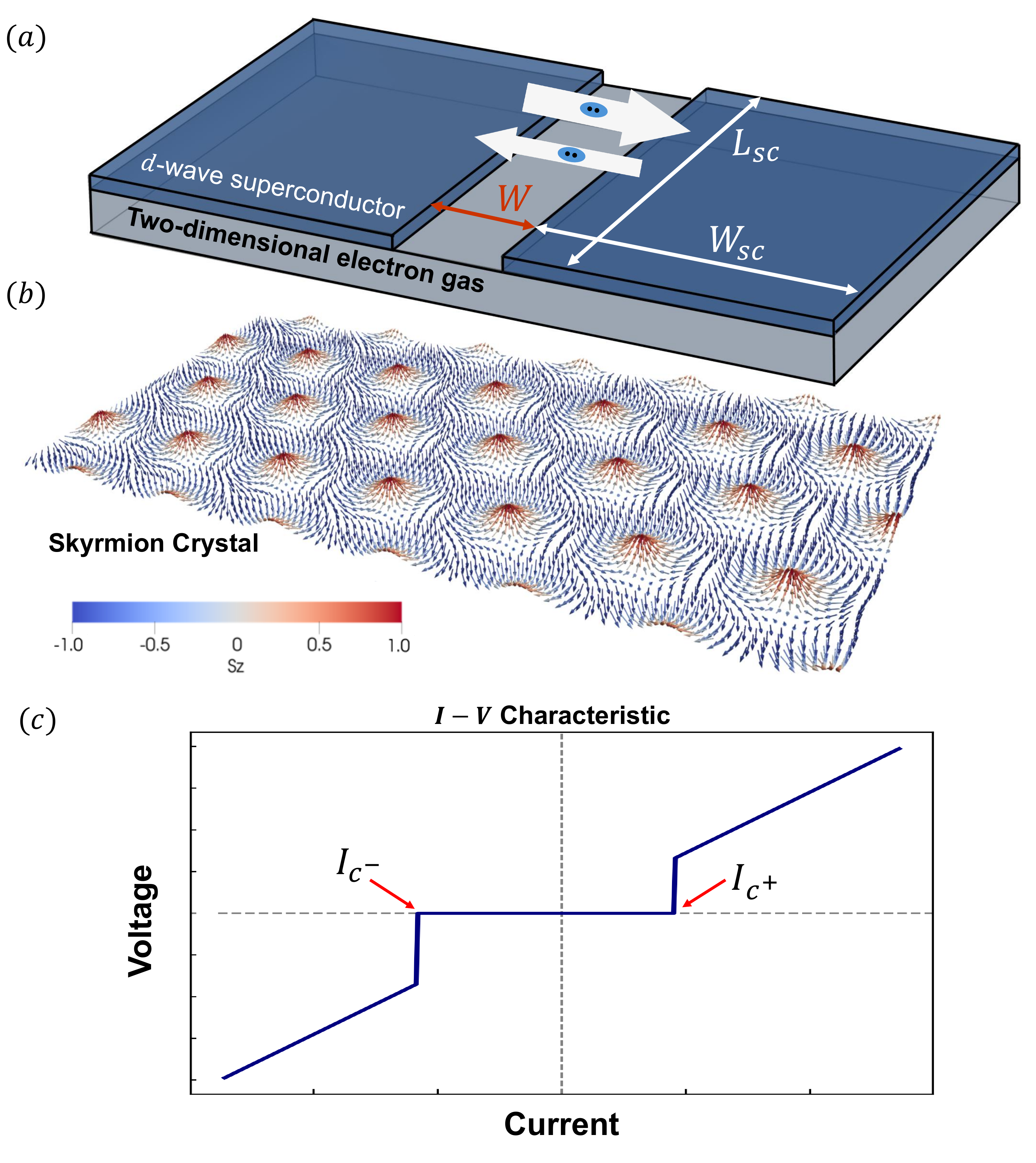}
\caption{(a) Schematic of the planar Josephson junction with two $d$-wave superconducting regions separated by a non-superconducting channel. (b) Spin configuration of a skyrmion crystal, which is placed underneath the planar Josephson junction. The arrows represent the $x$ and $y$ components of the spin vector ${\bf S}_i$, while the colorbar represents the $z$ component. (c) Representative current-voltage characteristic of the superconducting diode, showing asymmetry in the critical currents in forward and reverse directions.}
\label{FIG:1}
\end{figure}

Motivated by these developments, we investigate a planar Josephson junction consisting of two $d$--wave superconducting leads coupled through a two-dimensional electron gas (2DEG) that experiences both intrinsic Rashba spin--orbit coupling and an emergent Zeeman field from a skyrmion crystal. 
Our analysis focuses on how the interplay between  $d$--wave superconducting amplitude, and skyrmion-induced magnetic chirality gives rise to an anomalous Josephson phase shift and pronounced nonreciprocity in the current–phase relation. 
We demonstrate that the emergent real--space gauge field associated with the skyrmion texture provides a microscopic mechanism for symmetry breaking, resulting in large diode efficiencies. 
By mapping the microscopic current–phase relation onto macroscopic $I$--$V$ characteristic (Fig~\ref{FIG:1}(c)) via the resistively and capacitively shunted junction (RCSJ) model, we confirm diode--like behavior with strong gate tunability.

This study establishes a new microscopic route to realizing superconducting diodes with unconventional superconductivity by exploiting the topology of magnetic textures rather than externally applied fields. 

\section{Model and Method}

We consider a planar Josephson junction architecture, as depicted in Fig.~\ref{FIG:1}(a). 
The system comprises two $d$-wave superconducting leads that induce superconductivity via proximity effect into a confined 2DEG. 
The leads are separated by a normal (non-superconducting) region of width $W$. 
The entire 2DEG experiences intrinsic Rashba spin–orbit coupling due to structural inversion asymmetry and a spatially varying Zeeman field originating from an underlying skyrmion crystal, shown in Fig.~\ref{FIG:1}(b). 
This hybrid device architecture is experimentally feasible using 2DEG systems, combined with magnet–superconductor heterostructures. 

We employ a tight-binding Bogoliubov–de~Gennes (BdG) formalism to model this system microscopically. 
The Hamiltonian is given by:
\begin{align}
\mathcal{H} &= - t \sum_{\langle i j\rangle, \sigma} \left(c^\dagger_{i\sigma}c_{j\sigma} + {\rm H.c.}\right) + \sum_{i, \sigma} (4t - \mu) c^\dagger_{i\sigma}c_{i\sigma} \nonumber \\
&+ E_z \sum_{i, \sigma, \sigma^\prime} \left(\mathbf{S}_i \cdot \bm{\sigma}\right)_{\sigma \sigma^\prime} c^\dagger_{i\sigma} c_{i\sigma^\prime} 
+ \sum_{\langle i j\rangle} (\Delta_{ij} e^{i\varphi_{ij}} c^\dagger_{j \uparrow}c^\dagger_{i\downarrow} + {\rm H.c.}) \nonumber \\
&- i E_\alpha\sum_{\langle i j\rangle, \sigma \sigma^\prime} \left[(\bm{\sigma} \times \mathbf{d}_{ij})^z\right]_{\sigma \sigma^\prime} c^\dagger_{i\sigma} c_{j\sigma^\prime}, 
\label{eq:hamiltonian}
\end{align}
where $c^\dagger_{i\sigma}$ ($c_{i\sigma}$) creates (annihilates) an electron at lattice site $i$ with spin $\sigma \in \{\uparrow, \downarrow\}$. 
The first term describes nearest-neighbor hopping with energy $t = \hbar^2 / (2 m^* a^2)$, where $m^*$ is the effective electron mass and $a$ is the lattice spacing. 
The second term sets the chemical potential $\mu$, experimentally tunable via gate voltage. 
The third term represents Zeeman coupling, with strength $E_z$, between electron spin $\bm{\sigma}$ (Pauli matrices) and local magnetization $\mathbf{S}_i$ of the skyrmion texture. 
The fourth term describes proximity-induced $d_{x^2-y^2}$-wave superconductivity, with pairing amplitude $\Delta_{ij}$ non-zero only in regions under superconducting leads, and phase factor $\varphi_{ij}$ incorporating macroscopic phase difference $\varphi$ across the junction ($\varphi_{ij} = -\varphi/2$ for the left lead, $\varphi_{ij} = +\varphi/2$ for the right lead). 
The final term represents Rashba spin–orbit coupling with strength $E_\alpha$, where $\mathbf{d}_{ij}$ is the unit vector from site $i$ to $j$.

The skyrmion crystal texture is modeled as a periodic Néel-type configuration given by
$\mathbf{S}_i = S(\sin \theta_i \cos \phi_i, \sin \theta_i \sin \phi_i, \cos \theta_i)$, 
with angles as functions of lattice coordinates defining a periodic array of topological spin textures. The lattice grid spacing $a\!=\!10$~nm is used throughout the paper. The dimensions of the geometry are $W = 50~\text{nm},~ L_{sc} = 210~\text{nm}, \text{and}~ W_{sc} = 150~\text{nm}$. The radius of the skyrmions ($R_{Sk}$) used in the skyrmion crystal texture is $100~\text{nm}$.
The parameters used in the simulation are: $ t = 22.4~\text{meV}$,  $\Delta_0 = 4.0~\text{meV}$, and  $E_\alpha = 4.0~\text{meV}$ unless specified otherwise.

 The eigenvalues $E_n$ and eigenvectors $\psi_i^n \!=\! [u_{i\uparrow}^n, u_{i\downarrow}^n, v_{i\uparrow}^n, v_{i\downarrow}^n]^T$ of the Hamiltonian~(\ref{eq:hamiltonian}) were obtained by diagonalizing it using the unitary transformation ${c}_{i \sigma} = \sum_{n}u^n_{i \sigma}{\gamma}_n + v^{n *}_{i \sigma} \gamma^\dagger_n$, where $u^n_{i \sigma}$ ($ v^{n }_{i \sigma}$) represents quasi-particle (quasi-hole) amplitudes respectively, and ${\gamma}_n$ (${\gamma}_n^\dagger$) represents fermionic annihilation (creation) operator of the BdG quasi-particles corresponding to the $n^{\rm th}$ eigenstate~\cite{Sharma_Phys.Rev.B_2025}.
Further, the thermodynamic free energy at temperature $T$ is calculated as~\cite{Beenakker_1992, Beenakker_1992_QPC}
\begin{align}
    \mathcal{F}(\varphi) = -k_B T \sum_{E_n > 0} \ln\left[ 2 \cosh\left( \frac{E_n}{2k_B T} \right) \right].
\end{align}
 where $k_B$ is Boltzmann constant.
 
The Josephson supercurrent $I_s(\varphi)$ is then derived from the phase derivative of the free energy,
\begin{align}
    I_s(\varphi) = \frac{2e}{\hbar} \frac{d\mathcal{F}}{d\varphi}. 
\end{align}

\begin{figure}[t]
\includegraphics[width=0.9\linewidth]{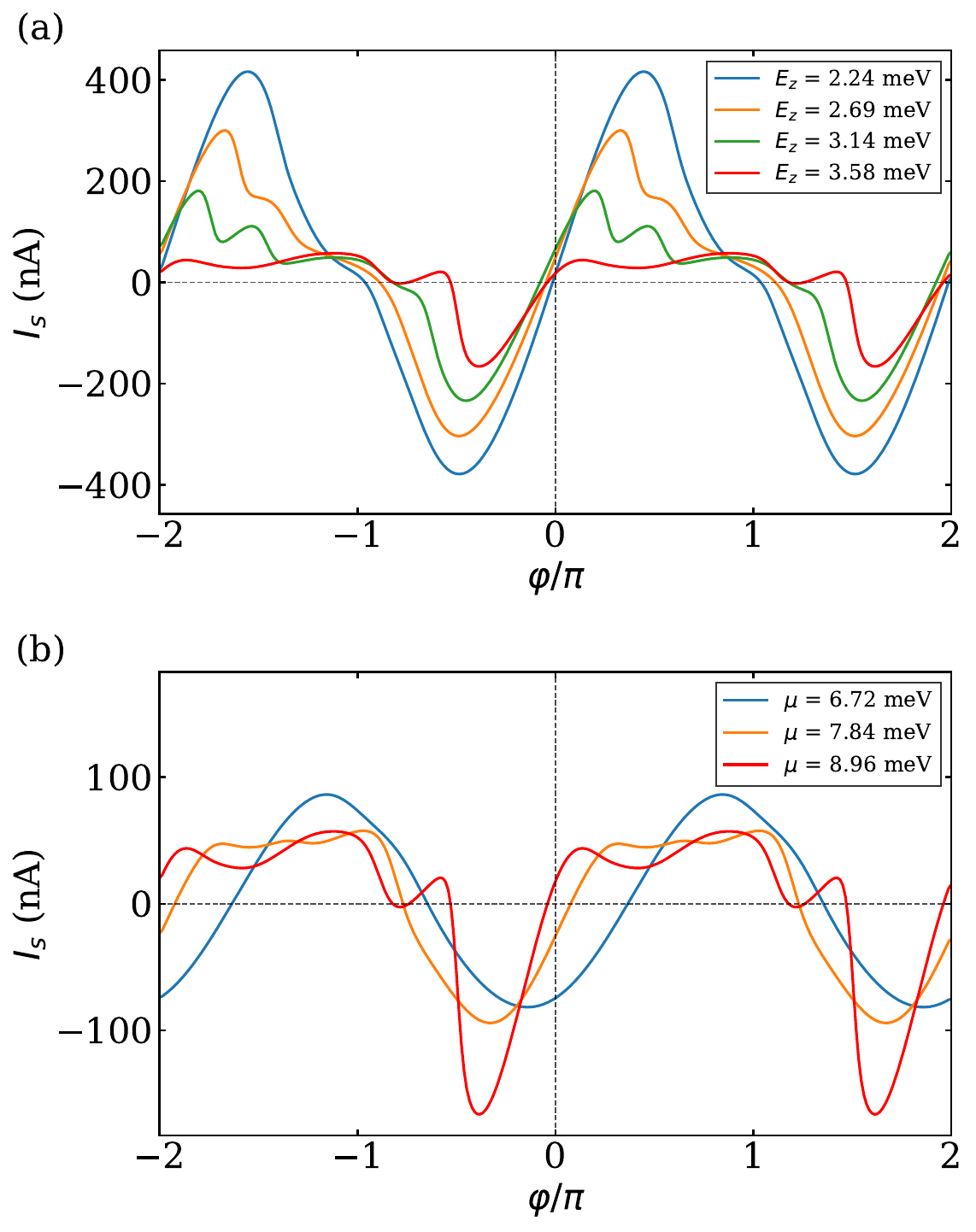}
\caption{(a) Calculated current--phase relations for different Zeeman couplings $E_z$ at fixed $\mu = 8.96~\text{meV}$. Increasing $E_z$ leads to stronger asymmetry and higher diode efficiency $\eta$. (b) Gate-tunable current–phase relations for varying chemical potential $\mu$ at fixed $E_z = 3.58~\text{meV}$.}
\label{FIG:2}
\end{figure}

This formulation captures the microscopic current--phase relation, which forms the basis for analyzing the diode response presented in the results section (Fig.~\ref{FIG:2}).

To connect the microscopic current-phase relation to measurable $I$--$V$ characteristics, we employ the RCSJ model. The numerical implementation of this model, including the dimensionless formulation and integration scheme, is detailed in Appendix~I. 
We can calculate the efficiency of the superconducting diode by following relation:
\begin{equation}
\eta = \frac{\left| I_c^{+} + I_c^{-} \right|}{\left| I_c^{+} \right| + \left| I_c^{-} \right|}
\end{equation}
where $I_c^{+}$ and $I_c^{-}$ denote the critical currents in forward and reverse bias directions, respectively. This definition ensures $\eta=0$ for a perfectly symmetric current-phase relation and $\eta=1$ for ideal diode-like behavior.

\begin{figure}[t]
\includegraphics[width=0.9\linewidth]{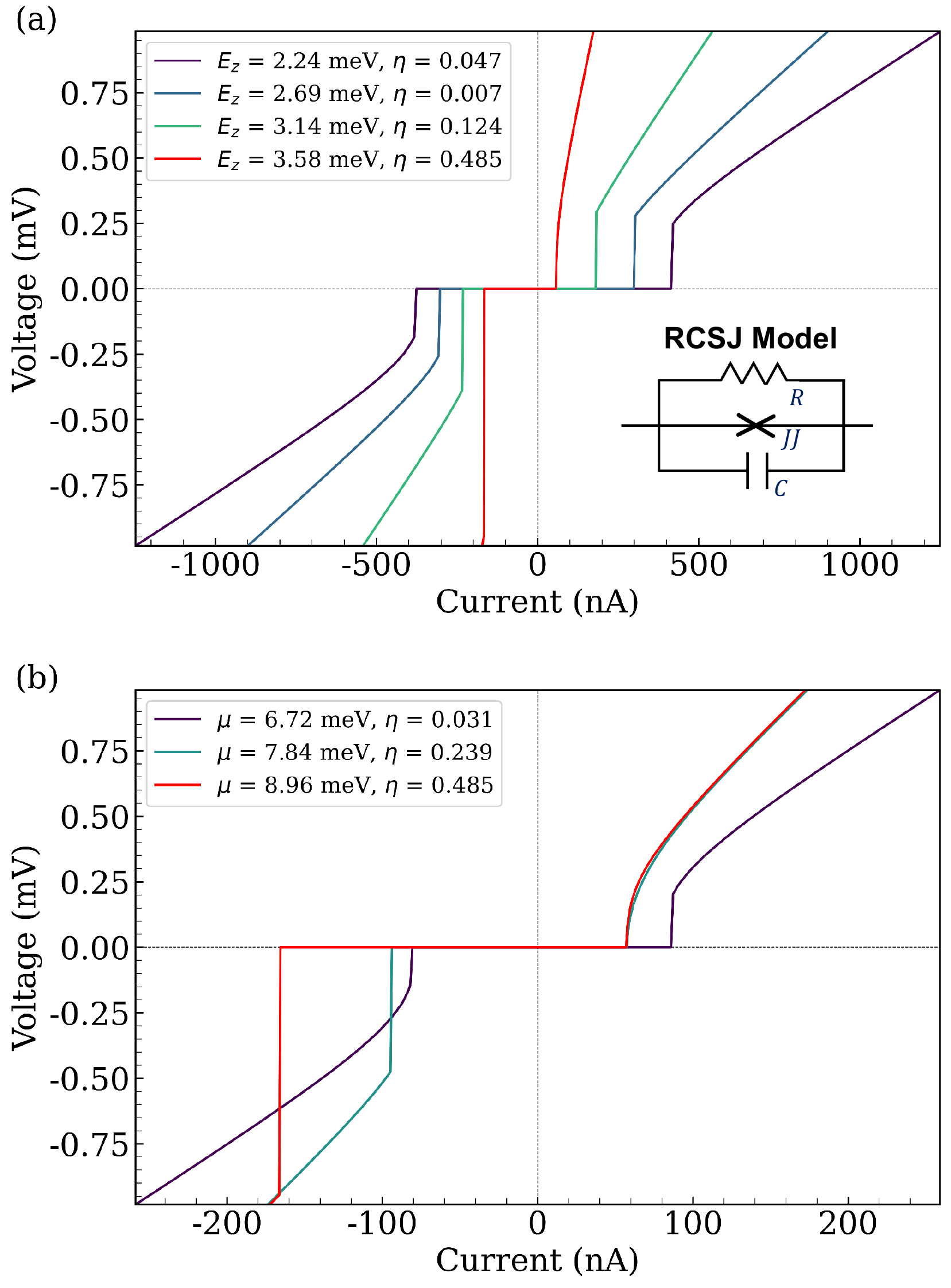}
\caption{Simulated current-voltage characteristics using the RCSJ model. (a) Variation with Zeeman coupling at fixed $\mu = 8.96~\text{meV}$. (b) Variation with chemical potential at fixed $E_z = 3.58~\text{meV}$. Asymmetric switching currents confirm diode operation. Inset in (a) shows a representative diagram for RCSJ model. }
\label{FIG:3}
\end{figure}

\begin{figure*}[t]
\includegraphics[width=0.85\linewidth]{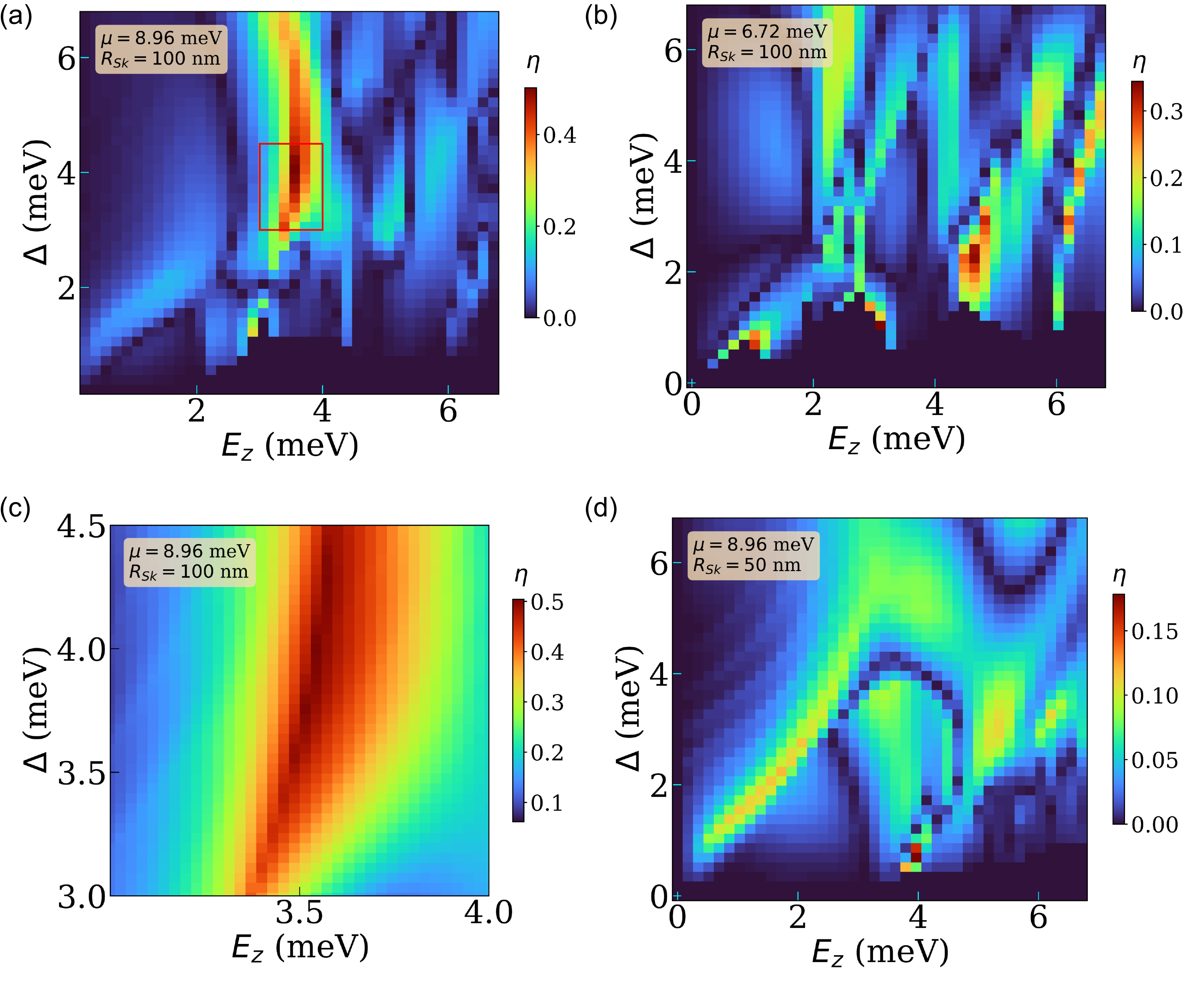}
\caption{ (a–d) Color maps of the diode efficiency $\eta$ as functions of Zeeman coupling $E_z$ and superconducting gap $\Delta$ at a fixed temperature $T = 0.1$~K. 
    (a) For $\mu = 8.96~\text{meV}$ and $R_{\text{Sk}} = 100~\text{nm}$, $\eta$ exhibits an enhancement around $E_z \approx 3.58~\text{meV}$ and $\Delta \approx 4.0~\text{meV}$, reaching values up to $0.5$. 
    The red rectangle highlights the region of maximal efficiency, which is magnified in panel (c) to show the high-$\eta$ domain. 
    (b) For a lower chemical potential $\mu = 6.72~\text{meV}$ (at the same of $R_{\text{Sk}}$), the pattern of $\eta$ becomes more intricate, and the overall magnitude decreases, indicating a weaker diode response at lower carrier density. 
    (d) For a smaller skyrmion radius $R_{\text{Sk}} = 50~\text{nm}$, the value of $\eta$ is reduced. }
\label{FIG:4}
\end{figure*}

\section{Results}
The current-phase relation of the Josephson junction exhibits significant tunability through both Zeeman coupling and electrostatic gating. As shown in Fig.~\ref{FIG:2}(a), the current–phase relation evolves from nearly sinusoidal behavior at low Zeeman coupling ($E_z$) to strongly non-sinusoidal characteristics at higher $E_z$ values. This progression indicates the emergence of higher harmonic components in the supercurrent, which is essential for achieving nonreciprocal transport ($|I_c^{+}| \neq |I_c^{-}|$), where $I_c^{\pm}$ denotes the critical currents in opposite current-flow directions. The temperature is kept fixed at $T = 0.1$~K throughout this paper. Also, the 2DEG can be electrostatically controlled via the chemical potential $\mu$. A variation in $\mu$ modifies the symmetry of the current–phase relation and introduces an anomalous phase shift, as shown in Fig.~\ref{FIG:2}(b). This anomalous phase shift enables a large nonreciprocal supercurrent flow through the junction. The current--voltage ($I$--$V$) characteristics, corresponding to the variations in $E_z$ and $\mu$, are shown in Figs.~\ref{FIG:3}(a) and \ref{FIG:3}(b), revealing large offsets between $I_c^{+}$ and $I_c^{-}$. Fig.~\ref{FIG:3}(a) shows the $I$--$V$ characteristic at different $E_z$ at a fixed value of the chemical potential $\mu=8.96~\text{meV}$. It is evident that this offset in the critical supercurrent increases with increasing the value of $E_z$. Figure~\ref{FIG:3}(b) shows the $I$--$V$ characteristic at fixed $E_z=3.58~\text{meV}$ for different values of $\mu$. The offset in the critical currents is tunable by controlling the value of $\mu$; the efficiency $\eta$ reaching a large value of approximately $49\%$. These results demonstrate that both Zeeman field and external gate potential can act as efficient control knobs for engineering superconducting diodes in the considered $d$-wave superconductor/skyrmion crystal hybrid platforms.

We varied also the superconducting pairing gap $\Delta$ to examine how the diode efficiency $\eta$ varies in Fig.~\ref{FIG:4} at different values of $\mu$, $E_z$ and $R_{\text{Sk}}$, the skyrmion radius. In Fig.~\ref{FIG:4}(a), we show the variation of $\eta$ in the plane of $E_z$ and $\Delta$, at a constant $\mu = 8.96~\text{meV}$ and $R_{\text{Sk}} = 100~\text{nm}$. In this parameter setting, $\eta$ is enhanced around $E_z \approx 3.58~\text{meV}$ and $\Delta \approx 4.0~\text{meV}$. Near this region, $\eta$ reaches large values up to approximately $0.5$, revealing a strong suppression of the critical supercurrent in one direction. The red rectangle in Fig.~\ref{FIG:4}(a) highlights this high-efficiency region, which is magnified in Fig.~\ref{FIG:4}(c) for clarity. Figure~\ref{FIG:4}(b) presents the corresponding efficiency for $\mu = 6.72~\text{meV}$ with the same skyrmion radius, $R_{\text{Sk}} = 100~\text{nm}$. In this case, the dependence of $\eta$ on $\Delta$ and $E_z$ becomes more intricate, and the magnitude of $\eta$ decreases on average. This reduction in $\eta$ indicates that a lower carrier density diminishes the strength of the diode response. Finally, Fig.~\ref{FIG:4}(d) shows the case of a smaller skyrmion radius, $R_{\text{Sk}} = 50~\text{nm}$. Despite the reduction in the overall magnitude of $\eta$, the parameter space over which it attains larger values becomes broader. This trend implies that the skyrmion radius $R_{\text{Sk}}$ can also control the diode efficiency in our proposed geometry significantly.



\section{Discussion}
The emergence of a robust nonreciprocal supercurrent transport in our skyrmion-based Josephson junction architecture can be understood through the interplay of multiple symmetry-breaking mechanisms. In our considered platform, the superconducting diode effect, with an efficiency approaching nearly 50\%, stems from the combination of $d$-wave superconducting pairing, Rashba spin--orbit coupling, and the topologically non-trivial magnetic texture.

Microscopically, the noncoplanar spin structure of the skyrmion crystal simultaneously breaks both inversion and time-reversal symmetries. The real-space gauge field, characterized by the skyrmion winding number, modifies the Andreev bound state spectrum such that $E_{\text{ABS}}(\varphi) \neq E_{\text{ABS}}(-\varphi)$, directly leading to the observed asymmetric current-phase relation where $|I_s(\varphi)| \neq |I_s(-\varphi)|$.

The $d$-wave order parameter amplifies this effect through its intrinsic sign change between crystallographic directions. The combination of $d$--wave pairing symmetry in the planar Josephson junction and the skyrmion crystal spin texture generates higher harmonic components in the supercurrent and a larger phase shift. This synergy between topology of real-space magnetic texture and momentum-space superconducting pairing anisotropy represents an unique attractive feature of our platform, which can be more advantageous compared to existing superconducting diode proposals.

We emphasize that our proposed platform can achieve large nonreciprocity without external magnetic fields, relying instead on the locally-varying magnetic texture of the skyrmion crystal. The observed gate-tunability via chemical potential $\mu$ further demonstrates electrical control on the diode efficiency, enabling reconfigurable superconducting circuit elements. The planar Josephson junction architecture further facilitates integration with existing fabrication processes and enables scaling towards more complex circuit geometries.

\section*{Acknowledgments}
PS acknowledges support from the Ministry of Education, India via a research fellowship. NM acknowledges support from Science and Engineering Research Board, India and SRIC office, IIT Roorkee (grant No. SRG/2023/001188 and IITR/SRIC/2116/FIG).

\section*{Appendix: Current-Voltage Calculation within the RCSJ Model}
The RCSJ \cite{inproceedings, Pegrum2023, articleRCSJ, Arnault_2021, Graziano_2020, Rangel_2016, Hens_2015, _onda_2015} model describes a Josephson junction as an ideal Josephson element shunted by a normal resistance \(R\) and a capacitance \(C\). The total current through the junction can be expressed as
\begin{equation}
I = I_c \sin\varphi + \frac{V}{R} + C\frac{dV}{dt},
\end{equation}
where \(\varphi(t)\) is the superconducting phase difference and \(V(t) = (\hbar/2e)\,\dot{\varphi}\) denotes the junction voltage. Substituting for \(V(t)\) yields
\begin{equation} 
I = I_c \sin\varphi + \frac{\hbar}{2eR}\dot{\varphi} + \frac{\hbar C}{2e}\ddot{\varphi},
\end{equation}
which can be rearranged as
\begin{equation}
\frac{\hbar C}{2e}\ddot{\varphi} + \frac{\hbar}{2eR}\dot{\varphi} + I_c \sin\varphi = I.
\label{eq:RCSJ_dim}
\end{equation}

For numerical convenience, Eq.~(\ref{eq:RCSJ_dim}) is cast into a dimensionless form by introducing normalized variables. The original equation can be rewritten as
\begin{equation}
I = I_s(\varphi) + \frac{\hbar}{2eR}\frac{d\varphi}{dt} + \frac{\hbar C}{2e}\frac{d^2\varphi}{dt^2},
\end{equation}
where \(I_s(\varphi)\) represents the supercurrent component.

\subsubsection{Normalization of Current}
Dividing both sides by the critical current \(I_c\) gives the normalized expression
\begin{equation}
\frac{I}{I_c} = \frac{I_s(\varphi)}{I_c} + \frac{\hbar}{2eRI_c}\frac{d\varphi}{dt} + \frac{\hbar C}{2eI_c}\frac{d^2\varphi}{dt^2}.
\end{equation}
Here, \(s(\varphi) = I_s(\varphi)/I_c\) defines the dimensionless current-phase relation.

\subsubsection{Normalization of Time}
Next, a dimensionless time variable \(\tau\) is introduced as
\begin{equation}
\tau = \frac{2eRI_c}{\hbar}t,
\end{equation}
which yields the following relations for time derivatives:
\begin{align}
\frac{d\varphi}{dt} &= \frac{2eRI_c}{\hbar}\frac{d\varphi}{d\tau}, \\
\frac{d^2\varphi}{dt^2} &= \left(\frac{2eRI_c}{\hbar}\right)^2 \frac{d^2\varphi}{d\tau^2}.
\end{align}

\subsubsection{Introduction of the Stewart–McCumber Parameter}
Substituting these derivatives back into the normalized equation gives
\begin{equation}
\frac{I}{I_c} = s(\varphi) + \frac{d\varphi}{d\tau} + \beta_c \frac{d^2\varphi}{d\tau^2},
\label{eq:RCSJ_2}
\end{equation}
where
\begin{equation}
\beta_c = \frac{2eR^2I_cC}{\hbar}
\end{equation}
is the dimensionless Stewart–McCumber parameter \cite{Araujo_Moreira_2005, Tr_as_2001, Hagenaars_1996, Hagenaars_1994}. This parameter quantifies the damping behavior of the junction: overdamped for \(\beta_c \ll 1\) and underdamped for \(\beta_c \gg 1\). The overdot now represents differentiation with respect to \(\tau\)
\cite{Trees_2005, Watanabe1997}.

Equation ~(\ref{eq:RCSJ_2}) is a second-order nonlinear differential equation. For numerical treatment, it is rewritten as a coupled system of first-order equations by defining the state vector
\begin{equation}
\vec{\zeta} =
\begin{bmatrix}
\varphi \\
\dot{\varphi}
\end{bmatrix},
\end{equation}
which leads to the following compact form:
\begin{equation}
\begin{bmatrix}
1 & 0 \\
0 & \beta_c
\end{bmatrix}
\dot{\vec{\zeta}} =
\begin{bmatrix}
0 & 1 \\
0 & -1
\end{bmatrix}
\vec{\zeta} +
\begin{bmatrix}
0 \\
I/I_c - s(\varphi)
\end{bmatrix}.
\label{dimeqn}
\end{equation}

The I–V characteristics are obtained by integrating this system for different bias currents \(I\). The numerical procedure proceeds as follows:

\begin{enumerate}
    \item \textit{Initialization:} The system is initialized at a given bias current with the state vector \(\vec{\zeta}(\tau=0) = (0,\,0)\), corresponding to zero phase and zero voltage.
    
    \item \textit{Time integration:} The equations are integrated over a finite interval \(\tau_{\text{span}}\) using the \texttt{solve\_ivp} function from the \textsc{SciPy} library, which employs an adaptive Runge–Kutta (RK45) algorithm \cite{Gammal_1999, Kondov_2001}. The current-phase relation \(s(\varphi)\) is implemented via cubic-spline interpolation of the theoretical data.
    
    \item \textit{Steady-state evaluation:} After transient dynamics subside, the steady-state phase velocity \(\langle \dot{\varphi} \rangle\) is evaluated as the time average over the latter portion of \(\tau_{\text{span}}\).
    
    \item \textit{Voltage determination:} The corresponding time-averaged DC voltage is obtained from
    \begin{equation}
    \langle V \rangle = \frac{\hbar}{2e}\left(\frac{2eRI_c}{\hbar}\right)\langle \dot{\varphi} \rangle = R I_c \langle \dot{\varphi} \rangle,
    \end{equation}
    or equivalently, using \(\beta_c = 2eR^2I_cC/\hbar\),
    \begin{equation}
    \langle V \rangle = \frac{\hbar\beta_c}{2eRC} \langle \dot{\varphi} \rangle.
    \end{equation}
    
    \item \textit{Current sweep:} The above procedure is repeated for successive current values \(I_n\). To ensure numerical continuity and reduce convergence time, the final state \(\vec{\zeta}\) from the previous current point \(I_n\) is used as the initial condition for the next, \(I_{n+1}\).
\end{enumerate}

We set $\beta_c=1$, yielding a critically damped regime. The value of RC is chosen equal to $10^{-12}$.

\bibliography{Ref}

\end{document}